# Delicate analysis of interacting proteins and their assemblies by flow field flow fractionation techniques

Aurélien Urbes[1,2,3], Marie-Hélène Morel[4], Laurence Ramos[1], Frédéric Violleau[2,3], Amélie Banc[1*]

[1] Laboratoire Charles Coulomb (L2C), Univ. Montpellier, CNRS, 34095 Montpellier, France.

[2] Laboratoire de Chimie Agro-industrielle LCA, Université de Toulouse, INRAE, INP-PURPAN, 31030 Toulouse, France.

[3] Plateforme TFFFC, Université de Toulouse, INP-PURPAN, 31076 Toulouse, France

[4] UMR IATE, Université de Montpellier, INRAE, Montpellier SupAgro, 2 pl. Pierre Viala, 34060 Montpellier, France.

* Email : amelie.banc@umontpellier.fr

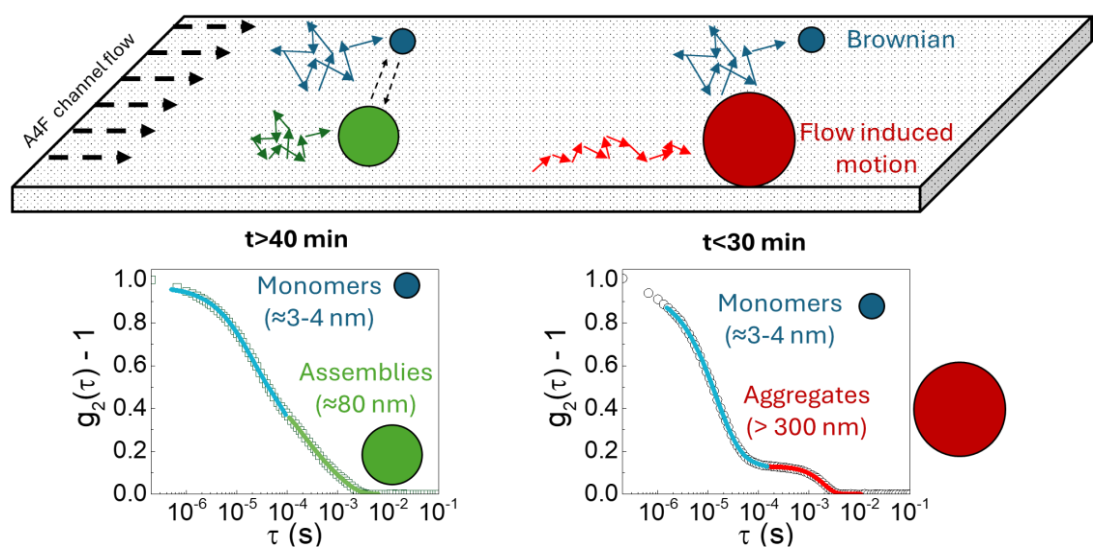

**Abstract:**

We study the efficiency of several Asymmetrical Flow Field-Flow Fractionation (AF4) techniques to investigate the self-associating wheat gluten proteins. We compare the use of a denaturing buffer including sodium dodecyl sulfate (SDS) and a mild chaotropic solvent, water/ethanol, as eluent, on a model gluten sample. Through a thorough analysis of the data obtained from coupled light scattering detectors, and with the identification of molecular composition of the eluted protein, we evidence co-elution events in several conditions. We show that the focus step used in conventional AF4 with the SDS buffer leads to the formation of aggregates that co-elute with monomeric proteins. By contrast, a frit-inlet device enables the fractionation of individual wheat proteins in the SDS buffer. Interestingly conventional AF4, using water/ethanol as eluent, is an effective method for fractionating gluten proteins and their complex dynamic assemblies which involve weak forces and are composed of both monomeric and polymeric proteins.

**Key words:** Flow Field Flow Fractionation, gluten, proteins, surfactant, supramolecular assemblies

**Introduction**

The characterization of supramolecular assemblies, such as protein complexes, protein micelles, protein aggregates, protein-polymer conjugates, is essential in biology, drug development and food processing. These assemblies are often governed by dynamic association equilibria that can be perturbated by the conditions of analysis. Indeed, depending on the technique, sample drying (e.g. electronic and atomic force microscopies in classical conditions), sample labelling (e.g. fluorescence microscopies), sample dilution (e.g. scattering techniques) or solvent change (e.g. for electrospray ionization in mass spectroscopy, for protein unfolding in SDS-PAGE or Size Exclusion Chromatography (SEC)) is required, challenging the structural stability of the assemblies. In addition, supramolecular objects can involve several different molecules, and separation techniques appear thus as techniques of choice to investigate their fluctuating internal composition. Chromatographic techniques are efficient to separate molecules but can denature supramolecular objects due to shear forces induced by the stationary phase[1,2]. By contrast, Asymmetrical Flow Field-Flow Fractionation (AF4) is a separation technique that reduces shear forces due to the absence of a solid phase in the channel. Retention is insured by a flow field perpendicular to the flow channel that concentrates species close to the bottom semi-permeable channel wall, the accumulation wall. The velocity of migration of species along the flow axis is controlled by the concentration profile perpendicular to the channel that is the result of the crossflow rate and the Brownian diffusion of the species. In addition, due to the use of an open channel, the accessible hydrodynamic size distribution ranges from nanometer to micrometer and is thus wider than with chromatographic techniques. Developments in AF4 for studies of the interactions between various systems, such as protein-protein, polymer-polymer, nanoparticle-drug, and nanoparticle-protein, are described in a recent review[3].

The wheat flours proteins, namely gluten, are important in the food industry because of their unique rheological, self-healing and gas barrier properties[4–6]. These properties are essential in the control of the texture of bakery products but present also great interest in the development of plant sourced meat analogues[7]. The origin and the variability of these properties are associated to the genetically controlled molecular protein composition but also to supramolecular structuration that derives from both biosynthesis and processing conditions (hydration, shear, temperature…). The structural analysis of these proteins is challenging due to gluten insolubility in water. The issue is circumvented by using buffers including surfactant in which protein-surfactant complexes are formed[8]. Furthermore, to ensure a total solubilization of gluten proteins in buffers, a sonication step is often used[9]. The molecular composition of gluten is characterized by a wide distribution of proteins that can be divided in two main classes[10,11]. Gliadin includes monomeric proteins, subdivided in α, β, γ and ω-gliadin, with molar masses comprised between 30 and 50 kg/mol, while glutenin refers to polymers of polypeptides, the glutenin subunits, crosslinked by intermolecular disulfide bonds, with a distribution of molar masses from 100 kg/mol to more than 1 000 kg/mol [12]. The highest molar masses of glutenin polymers are often ill defined due to the exclusion limit of SEC columns. That is why AF4 was initially identified as an interesting technique to characterize the size distribution of wheat proteins.

Wahlund and al.[13] and Stevenson et al.[15,16] were the first to investigate wheat flour protein extracts by AF4. In these early studies, the radius of gyration of the fractionated species was calculated from their retention time or using well known proteins as standards. However, it was shown that many experimental parameters, such as the quantity of the injected sample (volume and concentration), the crossflow rate, and the solvent quality could impact the retention time and direct size measurements by light scattering techniques were required[16]. Thus, the following studies used AF4 coupled with UV and Multi Angle Light Scattering (MALS) detectors[18–20] to properly determine sizes and masses. Very high molar mass components were identified in all studies from extracts rich in glutenin. However, molar masses (Mw) and radius of gyration (Rg) largely differs according to studies, with Mw ranging from 100 to $10^6$ kg/mol and Rg comprised between 40 and 80 nm. This variability can be attributed to many parameters including gluten protein extraction, eluent, AF4 method, light scattering data analysis. The first studies[13,14] used eluents comprising surfactants (anionic SDS or nonionic FL-70), while more recent studies use dilute acetic acid or ethanol/water as carrier fluids to avoid the formation of micelles and protein/surfactant complexes [19,20]. From a technical point of view, the presence of surfactant would induce a lower resolution and a lack of reproducibility in fractograms[15]. In addition, the analysis of light scattering data is more delicate in the presence of surfactants. Indeed, this technique requires the evaluation of the refractive index increment, which is strongly dependent on the protein/surfactant ratio within complexes. Furthermore, surfactants can modify interactions between proteins. As a consequence, we have recently proposed to use water/ethanol as a mild chaotropic solvent to investigate supramolecular assemblies in a model gluten extract characterized by a molecular composition in glutenin and gliadin comparable to that of native gluten[20]. In all these studies, independently of the experimental conditions, very large species were evidenced by AF4. They were assimilated to polymeric proteins by some authors[13–18], whereas others identified supramolecular assemblies including both monomeric and polymeric gluten proteins[20,21].

In this context, the aim of this paper is to compare AF4 of wheat gluten proteins performed in the classical denaturing solvent of wheat protein, 0.1M phosphate buffer pH 6.8 + 0.1% SDS ($P_{0.1}$), which is expected to disrupt the weak interactions stabilizing protein assemblies, and in a weak chaotropic solvent, water/ethanol 50/50 v/v (WE), in which supramolecular assemblies from [22]. The study is designed to vary independently solvent quality and fractionation technique, using the same wheat protein extract. A conventional AF4 and a frit-inlet AF4 methods are compared for solvent $P_{0.1}$, and two injection volumes are tested for each condition. The fractionation efficiency of the different methods is probed thanks to a thorough analysis of in-line MALS and Dynamic Light Scattering (DLS) signals in combination with an extemporaneous SEC analysis of the eluted proteins fractions. Finally, the composition of the large species fractionated in the two solvents is compared. The methodology of the study is summarized in Figure 1.

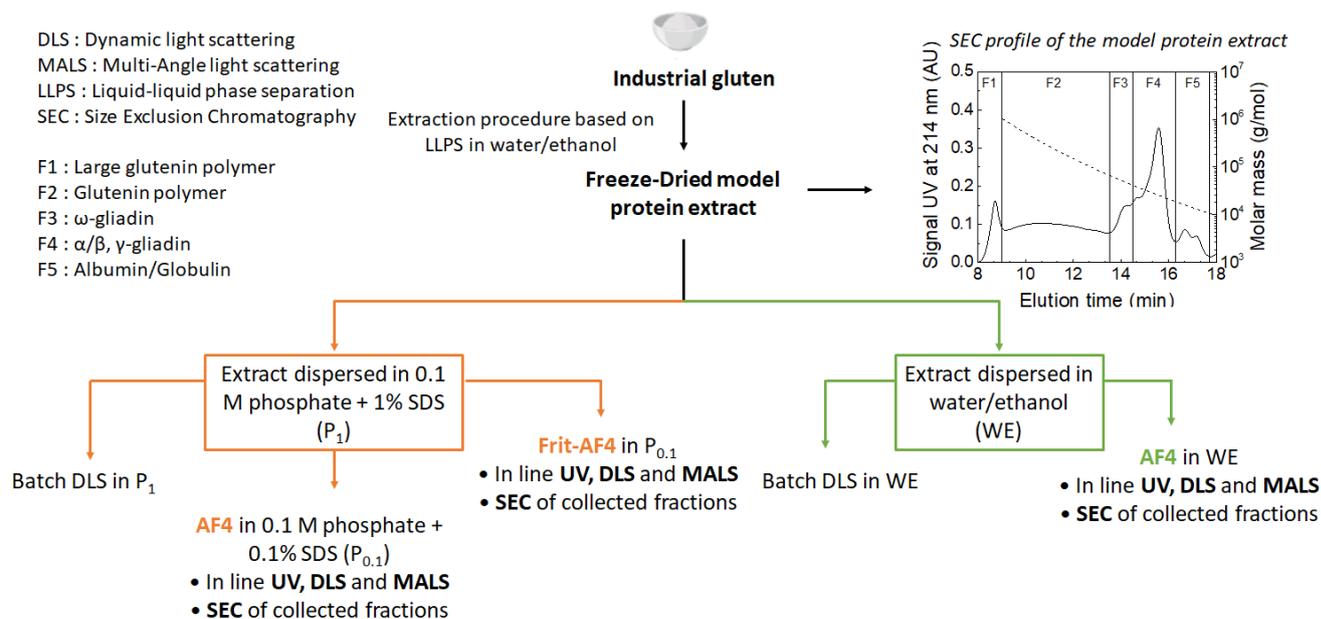

**Figure 1.** Methodology of the study. Note: Frit-AF4 was not performed in WE because the solvent viscosity leads to unmanageable back pressures at the flow rates required for frit-AF4.

1. **Materials and methods**

    1.1. **Materials**

Wheat protein is extracted from a commercial gluten (courtesy of Tereos-Syral Company). A "mild" protein extraction is performed in a mixture of deionized water and ethanol (purchased from Carlo Elba) (50/50, v/v) according to a protocol developed by Morel et al.[22]. Gluten (20 g) is added to 200 mL of the water/ethanol solvent, and mixed at 60 rpm for 19 h at 20°C. After centrifugation (30 min at 11 000 rpm), the supernatant (S1) is cooled to 6°C for 12 h to induce liquid-liquid phase separation. The dense phase (C2) is isolated after a 30 min centrifugation (11 000 rpm at 6°C). Five volumes of deionized water are added per volume of C2. The mixture is then frozen at -40°C and subsequently freeze-dried and powdered. The obtained model wheat protein isolate represents 17 % w/w of the initial gluten. It possesses an equilibrated amount of gliadin and glutenin, as checked by SEC analysis and is totally soluble in water/ethanol.

Two kinds of solvents are used to fully disperse the protein isolate: 0.1M phosphate buffer, pH 6.8, with 1% of sodium dodecyl sulfate (solvent $P_1$), and water/ethanol 50/50 v/v (solvent WE). Dispersions are prepared at 2 mg/mL and gently stirred during 2 hours at room temperature before analysis.

### 1.2. High-Performance Size Exclusion Liquid Chromatography (SEC)

SEC analysis of gluten proteins is carried out on a Dionex® Ultimate 3000 HPLC system equipped with a TSK G4000 SWXL column (Sigma-Aldrich) (30x7.8cm) preceded by a TSK 3000-SW guard column (Sigma-Aldrich) (4x6 cm). The flow rate is fixed at 0.7 mL/min with a 0.1 M sodium phosphate buffer at pH 6.8 + 0.1% of SDS (Fisher, France) (solvent $P_{0.1}$). Elution of protein samples (20 µL) prepared at 1 mg/mL in a denaturing buffer (0.1 M phosphate buffer pH 6.8+ 1% SDS, 6M urea) is recorded at 214 nm. The protein concentration is calculated from the UV signal using a value of 18.51 Lcm$^{-1}$g$^{-1}$ for the mass specific extinction coefficient[20]. The apparent molar mass calibration of the column is obtained using a series of protein standards with molar masses (Mw) in the range 13x10$^3$ - 2x10$^6$ g/mol according to Dahesh et al.[22]

### 1.3. Dynamic Light Scattering (DLS)

Batch dynamic light scattering experiments are performed with an Amtec goniometer at room temperature using scattering angles θ comprised between 30° and 130°, which correspond to wavevectors $q = \frac{4\pi n_o}{\lambda} \sin\left(\frac{\theta}{2}\right)$ (with $n_o$ the solvent refractive index and λ=663nm the laser beam wavelength) ranging from 8.12 to 29.50x10$^6$ m$^{-1}$ for solvent $P_1$, and 8.30 to 30.10x10$^6$ m$^{-1}$ for solvent WE. The auto-correlation functions are fitted with a double exponential function:

$$g_2(\tau) - 1 = [A_s . \exp(-\Gamma_s \tau) + A_f . \exp(-\Gamma_f \tau)]^2 \text{ (Eq.1)}$$

with the indexes *s* and *f* associated to slow and fast populations respectively. The decay rates $\Gamma_i$ are plotted as a function of q² to estimate the diffusion coefficients $D_i$ as $\Gamma_i = D_i q^2$. The hydrodynamic radii are calculated with the Stokes-Einstein equation, $R_i = \frac{k_b T}{6\pi \eta_o D_i}$, with $\eta_o$ the solvent viscosity ($\eta_o$ = 0.957 mPa s for $P_1$ and $\eta_o$ = 2.455 mPa s for WE at T=293K), $k_b$ the Boltzmann constant and T the temperature (K). The mass fraction of the slow population ($w_s$) is estimated from the relative amplitudes ($A_i$) and the size of the objects ($R_i$) according to the following equation [23,24]:

$$w_s = \lim_{q \to 0} \frac{\frac{A_s/(A_s+A_f)}{R_s^\beta}}{\frac{A_s/(A_s+A_f)}{R_s^\beta} + \frac{A_f/(A_s+A_f)}{R_f^\beta}} \quad (Eq. 2)$$

with β the Flory exponent that describes the solvent quality for polymeric objects. Here β=5/3 (associated to polymers in good solvent) is chosen for both populations of objects.

### 1.4. Asymmetrical Flow Field-Flow Fractionation (AF4)

Experiments are performed using a Dionex® Ultimate 3000 HPLC system coupled with an Eclipse AF4 system that regulates the flow into the channel during the fractionation. The separation occurs in a long asymmetrical channel whose dimensions are L = 26.5 cm, $b_0$ = 2.1 cm and $b_L$ = 0.6 cm with a spacer of 350 μm. We use a membrane made of regenerated cellulose with a molar mass cut-off of 10 kDa (Wyatt Technology). A 18 angles multi angle light scattering (MALS) DAWN HELEOS II apparatus (Wyatt Technology) with a dynamic light scattering (DLS) detector fixed at an angle θ =99° (q = 2.38x10⁷ m⁻¹ in $P_{0.1}$ and q = 2.47x10⁷ m⁻¹ in WE) is used for the online determination of molar masses and hydrodynamic radii. The wavelength of the laser is 663.8 nm and MALS data are fitted according to a Berry analysis which is more accurate for particles larger than 100 nm [25]:

$$\sqrt{\frac{KC}{R_\theta}} = \sqrt{\frac{1}{Mw} + \frac{16\pi^2}{3\lambda^2}\frac{1}{Mw} <R_g>^2 \sin^2(\frac{\theta}{2})} \quad \text{(Eq.3)}$$

K is the optical constant $\frac{4\pi^2 n_o^2 (\frac{dn}{dC})^2}{N_a \lambda^4}$ (with $N_a$ the Avogadro number), C the mass concentration of the sample, $R_\theta$ the Rayleigh ratio, Mw the molar mass, $R_g$ the radius of gyration and $\lambda$ the laser beam wavelength. In general, the quantity $\sqrt{\frac{KC}{R_\theta}}$ is plotted a function of $\sin^2\frac{\theta}{2}$ and the molar mass Mw is determined by the value of the ordinate extrapolated at the origin of the linear plot. The dn/dC values (with n is the refractive index of the protein solution at the mass concentration C) are determined experimentally using the Optilab refractive index device (Wyatt technology) for two model protein extracts prepared according to [26] and displaying glu/gli ratios of 0.8 and 1.6. The freeze-dried extracts are initially dissolved in $P_1$ at 2 g/L to saturate proteins with SDS and then the solvent is changed to $P_{0.1}$ using centrifugal filters with a 10 kDa cutoff (Amicon). Samples are finally diluted with $P_{0.1}$ to obtain a protein concentration range comprised between 0.02 and 2 g/L. An average value, with an error bar associated to the possible composition variation of the eluting species, is defined for each solvent. In $P_{0.1}$, dn/dC = 0.15±0.03 mL/mg, and in WE dn/dC = 0.169±0.012 mL/mg[22]. The extinction coefficient used to calculate the concentration during fractionation is 18.51 mL.mg$^{-1}$.cm$^{-1}$. MALS, DLS and UV data are analyzed with the Astra Software (version 8.1.0 Wyatt technology). Auto-correlation functions obtained in $P_{0.1}$ are fitted with a double exponential decay as in bulk (Eq. 1). Auto-correlation functions obtained in WE are fitted with a cumulant model (Eq.4) at short time (between $10^{-6}$ and $7\times10^{-4}$ s) and a compressed exponential (Eq.5) at long time (between $1\times10^{-3}$ and $2\times10^{-2}$ s):

$$g_2(\tau) - 1 = A_{cu} \exp[-2\Gamma_{cu}\tau + (K_2^2 \tau^2)] \quad \text{(Eq.4)}$$

$$g_2(\tau) - 1 = [A_{co} \exp(-(\Gamma_{co}\tau)^\beta)]^2 \quad \text{(Eq.5)}$$

With, $A_{cu}$ and $A_{co}$ the amplitudes, $K_2$ the second cumulant and $\beta>1$ the compression factor. For the cumulant analysis, the apparent hydrodynamic radii are estimated using $R_{cu} = \frac{k_b T q^2}{6\pi \eta_o \Gamma_{cu}}$ and polydispersities are given by $\sigma = \frac{K_2}{\Gamma_{cu}}$.

To determine the void volume between the different detectors and to normalize the response of the photodiodes of the MALS detector, a 2 g/L bovine serum albumin (BSA) solution is used. To get reproducible results, several injections of the protein model extract dispersed in $P_1$ or WE are performed before the analysis of the fractogram. 50 µL or 200 µL of protein dispersions at 2 g/L are injected for all experimental conditions.

### 1.4.1. AF4 conditions

Figure 2a displays the AF4 method used for the analysis. Sample injection starts at time t=3 min for a duration of 3 min with a flow rate of 0.2 mL/min. The focus step starts at t=2 min and lasts 6 min with a crossflow rate of 1.5 mL/min. At the end of the focus step, the elution step starts with an isoforce crossflow rate fixed at 1.5 mL/min during 20 min allowing the separation of the smallest objects. To facilitate the elution of the largest objects, a linear decrease of the crossflow from 1.5 mL/min to 0.10 mL/min in 22 min is imposed. Then, the crossflow is maintained at 0.10 mL/min during 10 min. Once the crossflow is stopped, the injection loop is washed during 5 min. The detector outlet flow is fixed at 0.6 mL/min to accommodate for the high viscosity of the WE solvent ($\eta_o$=2.455 mPa s at 25°C). The same crossflow program is used with $P_{0.1}$ as eluent to assess the impact of the eluent nature on protein fractionation.

### 1.4.2. Frit-AF4 conditions

The method used for the Frit-AF4 analysis is displayed in Figure 2b. Unlike in a conventional AF4 procedure, there is no focus step. In frit-inlet method, the sample is injected with a stop-less flow, and the hydrodynamic relaxation is achieved thanks to the introduction of the sample into the frit area where a fast crossflow rate drives the sample to the wall of the channel. Here, the sample injection flow rate of the sample is fixed at 0.2 mL/min with an isoforce frit-inlet flow rate fixed at 4 mL/min during 20 min. The outlet flow rate is fixed at 1 mL/min. The ratio between the sample flow and the frit-inlet rate is 0.05. This ratio remains within the range determined by Moon and al. [27] as sufficient to lead to a distribution of particles at equilibrium in the channel cross section. After 11 min of constant crossflow, the crossflow decreases linearly from 4 mL/min to 0.10 mL/min during 20 min. The sample injection flow is then stopped, and the crossflow is maintained at 0.1 mL/min during 20 min before being set to 0. Finally, the injection loop is washed during 5 min.

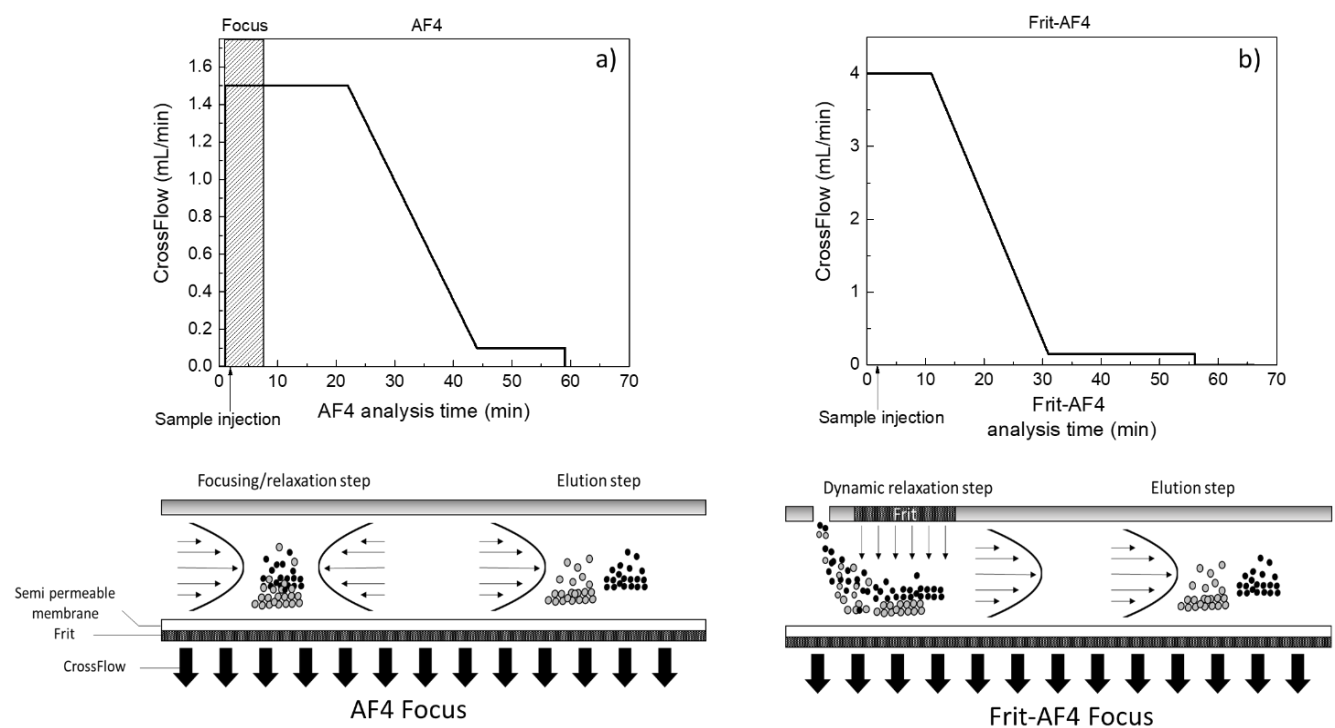

**Figure 2.** Crossflow programs for (a) AF4 method in 0.1M phosphate, 0.1% SDS ($P_{0.1}$) and water/ethanol 50/50 v/v (WE). and for (b) Frit-AF4 method in $P_{0.1}$. The dashed zone corresponds to the focusing step in (a).

### 2.4.3 Collection of fractions

We collect the eluted flows with a binning time of 5 minutes for AF4 (6 to 8 collected samples), and of 3 minutes for Frit-AF4 (14 collected samples) using the Fra-920 collector (Amersham Biosciences). The total volume of sample injected is 200 μL. The fractionation is repeated two to three times for each method, and collected samples are pooled on the basis of their elution times. For SEC analysis, collected and pooled samples are concentrated 25 times using centrifugal filters (Amicon) with a 10 kDa molecular weight cutoff while being equilibrated in sodium phosphate buffer (pH 6.8) including 1% SDS and 6M urea.

## 2. Results
### 2.1. Batch analysis of the model gluten isolate

The protein composition of the model gluten proteins extract is characterized by a well-established SEC method whose capability to fractionate glutenin polymers from monomeric gliadins was confirmed by SDS-PAGE analysis of the eluted SEC fractions [28,29]. Samples are firstly diluted in a strongly denaturing buffer comprising 1% SDS and 6M urea to preliminary destroy all non-covalent intermolecular interactions and unfold proteins through SDS saturation, and then eluted with a 0.1% SDS buffer on a SEC column. The SDS content in the denaturing buffer is high enough to saturate proteins with the surfactant (saturation is expected at about 1.7 g SDS/g protein) [30,31], while the SDS concentration in the elution buffer is ten times less to limit the formation of free additional SDS micelles. The SEC profile of the extract used for the study is illustrated in Figure 1. Between elution times of 8 min and 9 min, proteins are excluded from the pores of the stationary phase and thereby eluted at the column void volume, leading to an exclusion peak. These proteins are categorized as large molecular weight glutenin polymers (HMW-GP) (Fraction F1). The fraction F2, which elutes between 9.5 and 13.5 min, corresponds to glutenin polymers with a molar mass comprised between $10^5$ and $10^6$ g/mol. Proteins eluted between 13.5 and 16.25 min are the various monomeric gliadins; fraction F3 corresponds to ω-gliadins, while the α, β and γ-gliadins contribute to the F4 fraction. The fraction between 16.25 and 17 min includes traces of the water and salt soluble albumins and globulins that have been partly solubilized in water/ethanol. The proteins extracted from the gluten powder have a high percentage of glutenin polymers, 46%, which corresponds to a glutenin/gliadin mass ratio, Glu/Gli = $\frac{F_1+F_2}{F_3+F_4}$ =1.07.

The model gluten protein isolate is suspended at 2 g/L in 0.1 M sodium phosphate buffer (pH 6.8) + 1% SDS (solvent $P_1$) or water/ethanol (50/50, v/v) (solvent WE). The impact of the type of solvent on the size distribution of the proteins is studied using batch multiangle dynamic light scattering (DLS). The auto-correlation functions are poorly fitted by a single exponential decay function. Instead, a double exponential decay provides a good fit of the experimental data. An estimate of the sizes of two main populations is obtained from the fit parameters (See fitted data in figure SI1 of Supporting Information). The fast decay rate corresponds to the small objects, while the slow decay rate is associated with the large objects present in the sample. For both solvents, a linear dependance of the fast decay rate as a function of $q^2$ is measured in the whole q range, demonstrating a diffusive behavior. The hydrodynamic size calculated for the small objects is approximately 20 nm for both solvents (Table 1). The slow decay rates allow one to estimate the size of the large objects which we consider as supramolecular objects. In $P_1$, the hydrodynamic radius is about 245 ± 8 nm, compared to 129 ± 5 nm in WE. It is important to note that at high $q^2$ values, the slow decay rates does not increase linearly with $q^2$, possibly due to internal dynamics of the largest objects[32]. The weight fraction of assemblies is estimated at 7% in $P_1$ and 13% in WE.

To better describe the size distribution of wheat proteins and their assemblies in solvents $P_1$ and WE, the protein samples are fractionated using AF4.

| Solvent | $R_{hf}$ (nm) | $R_{hs}$ (nm) | $w_s$ (%) |
|---|---|---|---|
| $P_1$ - 0.1 M Phosphate + 1% SDS | 20 ± 1 | 245 ± 8 | 7.1 |
| WE - Water-Ethanol | 22 ± 1 | 129 ± 5 | 13.2 |

**Table 1.** Hydrodynamic radii and weight percentage of the population of large objects from batch multiangle DLS analysis

### 2.2. Fractionation in 0.1M phosphate buffer, pH 6.8 + 0.1 % SDS

#### 2.2.1. Conventional AF4

The gluten protein isolate is first suspended in buffer $P_1$ to ensure total saturation of proteins with SDS and subsequently fractionated by AF4 using solvent $P_{0.1}$, a phosphate buffer including only 0.1 % SDS, like in the SEC analysis. The SDS concentration in the eluent is close to the SDS critical micellar concentration. Hence, the elution solvent is expected to maintain SDS saturation of the proteins, while minimizing the number of free surfactant micelles. Figure 3a displays the results obtained with an injection volume of 200 µL. We find that protein elution begins at t=7 min just at the end of the focus step. A large protein concentration peak is eluted between t=7 and 25 min and then tails until 40 min. After 42 min, the time corresponding to the change of crossflow slope in the protocol (Figure 2a), another concentration peak, although quite small, is eluted until 55 min. Just after the focus step, between 7 and 13 minutes of elution, an unexpected drop of the apparent molar mass from $2.0 \times 10^6$ to $4.0 \times 10^5$ g/mol is recorded. The apparent molar mass starts to increase at 13 min and reaches a value of $1.2 \times 10^6$ g/mol at 25 min, and remains constant until 40 min. Finally, when the second concentration peak is eluted, between 40 to 55 min, the apparent molar mass drastically increases up to $3 \times 10^7$ g/mol.

The autocorrelation functions, calculated from the in-line DLS device show two clear characteristic decay times, around $10^{-5}$ s and $10^{-3}$ s, for all elution times. We find that the relative amplitudes of the decays significantly vary with elution time (Figure 3c). A double exponential model is initially used to fit the autocorrelation functions. The apparent hydrodynamic radii associated to the two characteristic times and the relative amplitude of the slow decay rate are plotted in Figure 3c. The smallest size is constant with the elution time, and its value, 3 nm, is consistent with the mean radius of gliadin[33]. The largest size would be associated to supramolecular objects. However, the size extracted from a double exponential decay (Eq.1) is not very reliable, because of the poor fit of the data. At long time, we find that the decay of the correlation functions is sharper than a simple exponential function. Interestingly, a compressed exponential function (Eq.5) with a compression factor β=2 (inset of Figure 3e) nicely fit the experimental data. Such compressed exponentials are typically measured for ballistic dynamics [34–36]. Here, they can be associated to the flowing motion of large scattering objects across the measuring cell which is faster than their diffusive dynamics. Therefore, the apparent radii estimated before 40 min are incorrect (empty squares in Figure. 3c) and represents only a lower cutoff size for the scattering objects. By contrast, the long time decay of the auto-correlation functions measured between 43 and 47 min is more stretched (β=0.7) and can be associated to the diffusive dynamics of polydisperse systems. The rough estimation of the size (full dots in Figure 3c), around 160 nm, is consistent, although slightly smaller, with the large size measured by batch DLS (239 nm). The relative amplitude of the slow decay, associated to the largest objects, progressively increases from 20 to 70% with the elution time. In short, in line DLS data clearly show that the sample is not well size fractionated as co-elution of small and very large objects is evidenced, which prevents any quantitative analysis of the molecular weight that is based on the scattering data of monodisperse samples. Similar conclusions are obtained with an injection volume of 50 μL (See Figures SI2 and SI3 in Supporting Information).

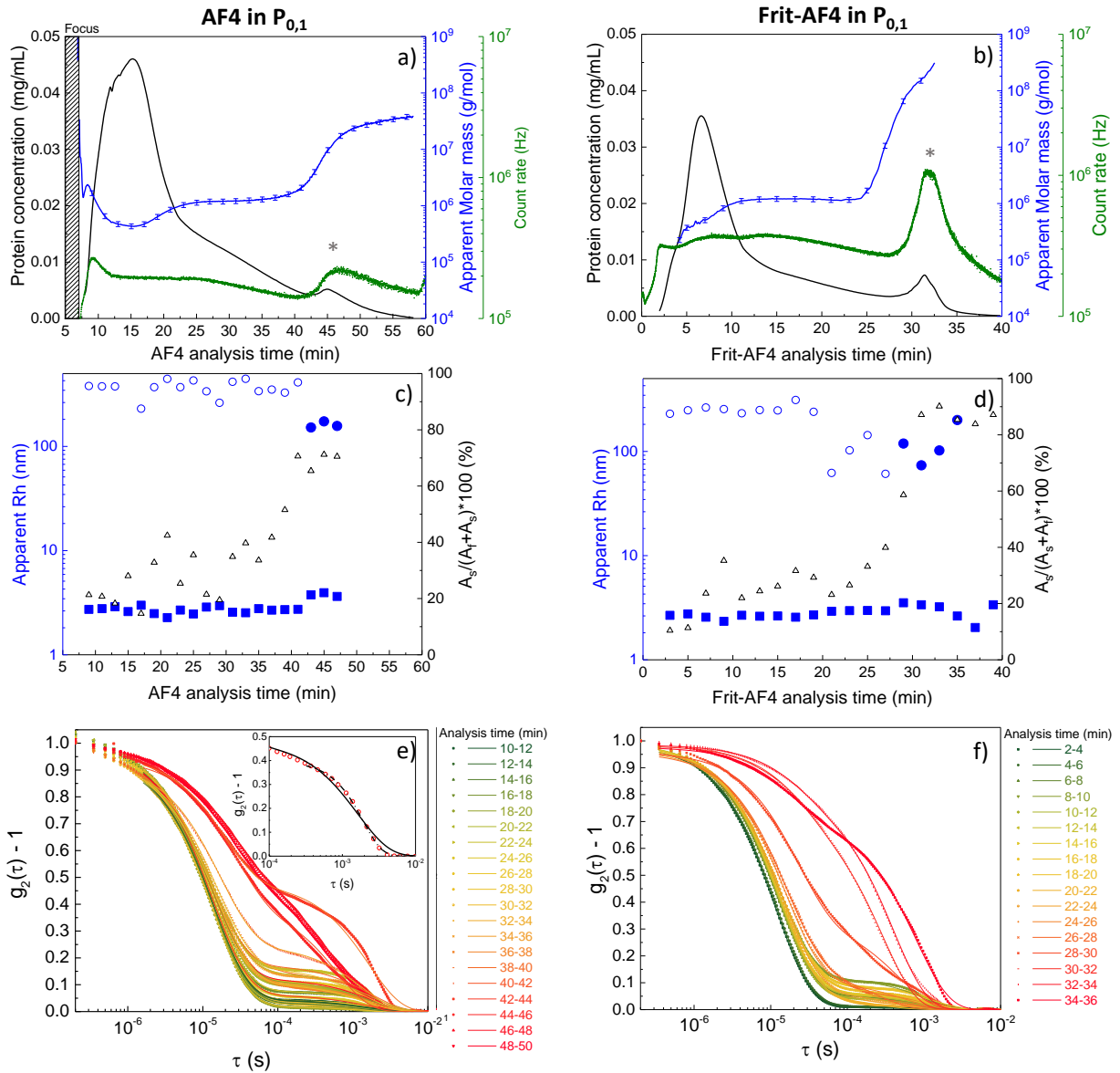

**Figure 3.** AF4 (Left column) and Frit-AF4 (right column) analysis in 0.1M phosphate, 0.1% SDS ($P_{0.1}$) (the injection volume is 200 µL). (a,b) Protein concentration (black line), apparent molar mass (blue line) and DLS count rate (green line) as a function of the AF4 analysis time. The grey star corresponds to the maximum of the MALS signals and the dashed area corresponds to the focus step. (c,d) Apparent hydrodynamic radii associated to the fast (blue squares) and the slow decay rates (blue circles) as function of the AF4 analysis time. Empty circles give the lower cutoff size of the large co-eluted objects at short analysis time. Empty triangles correspond to the amplitude percentage of the slow decay. (e,f) Auto-correlation

functions measured in-line along elution and averaged each 2 min. Inset of e) Zoom on the long delay auto-correlation function measured between 42 and 44 min. Symbols are experimental data and the full (resp. dashed) line is a fit with an exponential (resp. compressed exponential) function.

To identify the proteins eluted during the AF4 fractionation, 8 fractions cumulating elution periods of 5 minutes are successively collected, concentrated and analyzed by SEC. On the basis of the UV signal recorded during AF4, a protein recovery yield of 87±5% of the injected protein is calculated, indicating that only a small fraction of the injected protein is lost in the system (see Figure Table SI3 in Supporting Information). We compare on a mass basis, the SEC profile of the injected proteins adjusted for a global recovery of 87% to the one calculated from the cumulated SEC profiles of the 8 collected fractions. Both profiles are rather similar at the exception of fraction F1, which almost disappears (inset Figure 4a). The comparison reveals that all the proteins classes are lost in the fractionation system more or less to a same extent, at the exception of the largest glutenin polymers from F1. Despite an enrichment in glutenin at long AF4 analysis time, the normalized SEC profiles of the collection series are weakly contrasted (Figure 4a). Hence, protein fractionation remains limited, as already anticipated from in-line DLS analysis. Figure 4c displays how the content of each of the different gluten protein classes is divided along AF4 fractionation. At short analysis time (t<15 min), almost 30% of the injected glutenin polymers are eluted together with gliadins while the apparent molar mass is decreasing (Figure 3a). The fractions collected between 10 to 20 min are the richest in gliadin and the later collected samples display a quite similar protein composition despite a large change of apparent molar mass at 45 min. The poor fractionation is reflected in a very noisy evolution of the glutenin over gliadin ratio (inset Figure 4c).

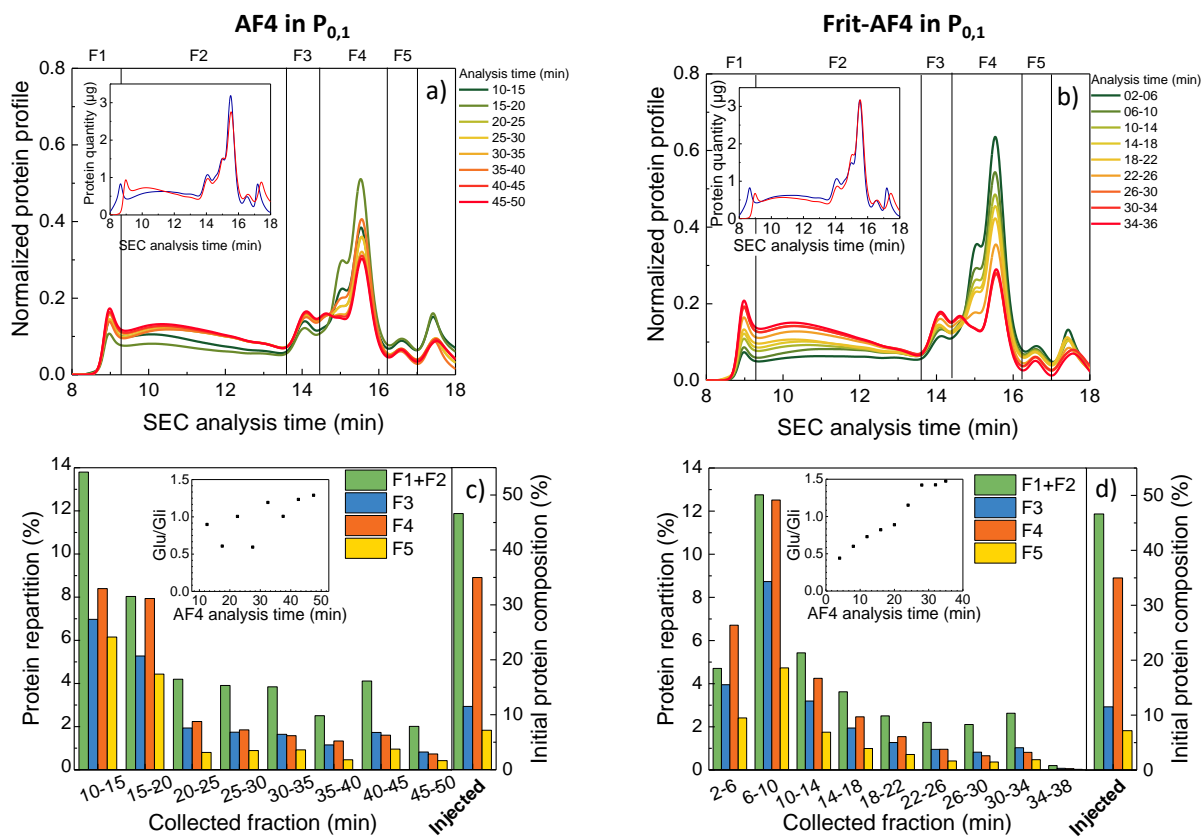

**Figure 4.** SEC analysis of the composition of collected samples from AF4 (Left column) and Frit-AF4 (right column) in 0.1M phosphate, 0.1% SDS ($P_{0.1}$) performed with an injection volume of 200 µL. (a,b) Normalized protein profiles (the total area is equal to 1) of collected samples. Inset: SEC profiles (expressed in mass) of the injected sample (blue) (adjusted for the global AF4 protein recovery yield calculated from the AF4 UV signal) compared to the sum of the SEC profiles of the collected protein fractions (red). (c,d) Distribution of the different classes of wheat proteins within collected fractions obtained along analysis and comparison with the injected sample composition. Green : glutenin polymers , blue : ω-gliadin, orange : α/β,γ-gliadin and yellow : Albumin/globulin. Inset : Evolution of the glutenin/ gliadin ratio as a function of the AF4 analysis time.

As a conclusion, the conventional AF4 fractionation of gluten proteins in $P_{0.1}$ is not efficient whatever the injection volume (200 and 50 µL) and AF4 has to be used very cautiously for the analysis of wheat proteins composition and propensity to assemble in supramolecular objects.

### 2.2.2. Frit-inlet AF4

To investigate whether the co-elution phenomenon observed in AF4 is induced by the focus step, we test a frit-inlet channel. This device avoids sample concentration in a thin layer, which can potentially contribute to protein aggregation. In the frit-inlet system, relaxation is achieved during the injection of the sample into the channel while the proteins are pushed towards the membrane by a flow passing through an upper frit. Hence, the proteins reach hydrodynamically their steady state equilibrium within the frit-inlet region[27].

Figure 3b displays the Frit-AF4 elution profile of the model gluten extract suspended in $P_1$ and eluted with $P_{0.1}$ using an injection volume of 200 µL. A large protein concentration peak is eluted between 2 min and 12 min, and then tails until 30 min. Beyond 30 min, corresponding to the change of imposed crossflow rate (Figure 2a), a smaller peak is eluted until 40 min. From 4 to 10 min, the apparent molar mass increases from $2 \times 10^5$ to $10^6$ g/mol and then keeps a constant value until 24 min. Then, a sharp increase of the apparent molar mass from $10^6$ to $10^8$ g/mol is measured before the apparition of the second concentration peak. The proportion of objects with an apparent mass greater than $10^6$ g/mol is comprised between 7 and 8 % of the total amount of protein, which is close to the estimation of the weight content of the slow population measured by batch DLS. By contrast with the conventional AF4, there is no decrease in the molar mass at the beginning of the fractogram. This suggests that co-elution is reduced with the frit-inlet system.

In-line DLS auto-correlation functions display two characteristic times as previously recorded during conventional AF4 (Figure 3f). The shorter time, $2.2 \times 10^{-5}$ s, is constant during the elution. It corresponds to hydrodynamic radii of the order $2.5 \pm 1$ nm, a numerical value comparable to the one measured in conventional AF4. Before 25 minutes, the longest decay time is well fitted with a compressed exponential decay. We find a characteristic time 1.5 times shorter than in conventional AF4. This is connected to the faster flow rate used in the frit-inlet method (1 mL/min instead of 0.6 mL/min). This confirms the flow-induced origin of this characteristic time. After 25 minutes of AF4 analysis time, long time decays are exponential, as expected for diffusive motion, and hydrodynamic radii of about 100 nm are measured. The relative amplitude of the long decay time increases with analysis time. In comparison with conventional AF4 in $P_{0.1}$, the relative amplitude of the largest objects is reduced at short time and more important at long time, which indicates a better size fractionation by the frit-inlet method.

This improved fractionation is confirmed by the SEC analysis of the collected fractions. Figure 4b shows that, while the global protein recovery is similar to the one obtained in conventional AF4 ($81 \pm 6\%$, see Supporting Information), a clear progressive evolution of the normalized protein profiles of collected fractions as the elution progresses is obtained. At short analysis times, samples are enriched in α/β, γ-gliadin (F4), while at long analysis times, samples are enriched in glutenin polymers (F1, F2) and ω-gliadin (F3). A progressive increase of the glutenin/gliadin ratio is observed in contrast with the conventional AF4 analysis (insets Figure 4c and 4d). Nevertheless, the fractions always include mixtures of all protein types (Figure 4b). In addition, no clear change of protein composition is associated with the large increase of apparent molecular weight at t=26 min. The frit-AF4 analysis was also performed with an injection volume of 50 μL (See Figures SI2 and SI3 in supporting information). In that case, co-elution signatures evidenced by DLS, are nearly totally suppressed and hydrodynamic radii, about 3 nm, are consistent with individual proteins. In addition, molar masses measured at short elution times (about 4-6x10$^4$ g/mol), are consistent with gliadin, while molar masses measured at longer times (2-8x10$^5$ g/mol) are consistent with glutenin polymers. Interestingly, no assemblies are fractionated while the protein recovery yield is good (87%), which suggests a full dissolution of assemblies in these conditions.

## 2.3. Conventional AF4 in ethanol/water

The wheat protein extract is solubilized and fractionated in a weak chaotropic solvent, water/ethanol (WE), using the same conventional AF4 method as the one used in P$_{0.1}$.

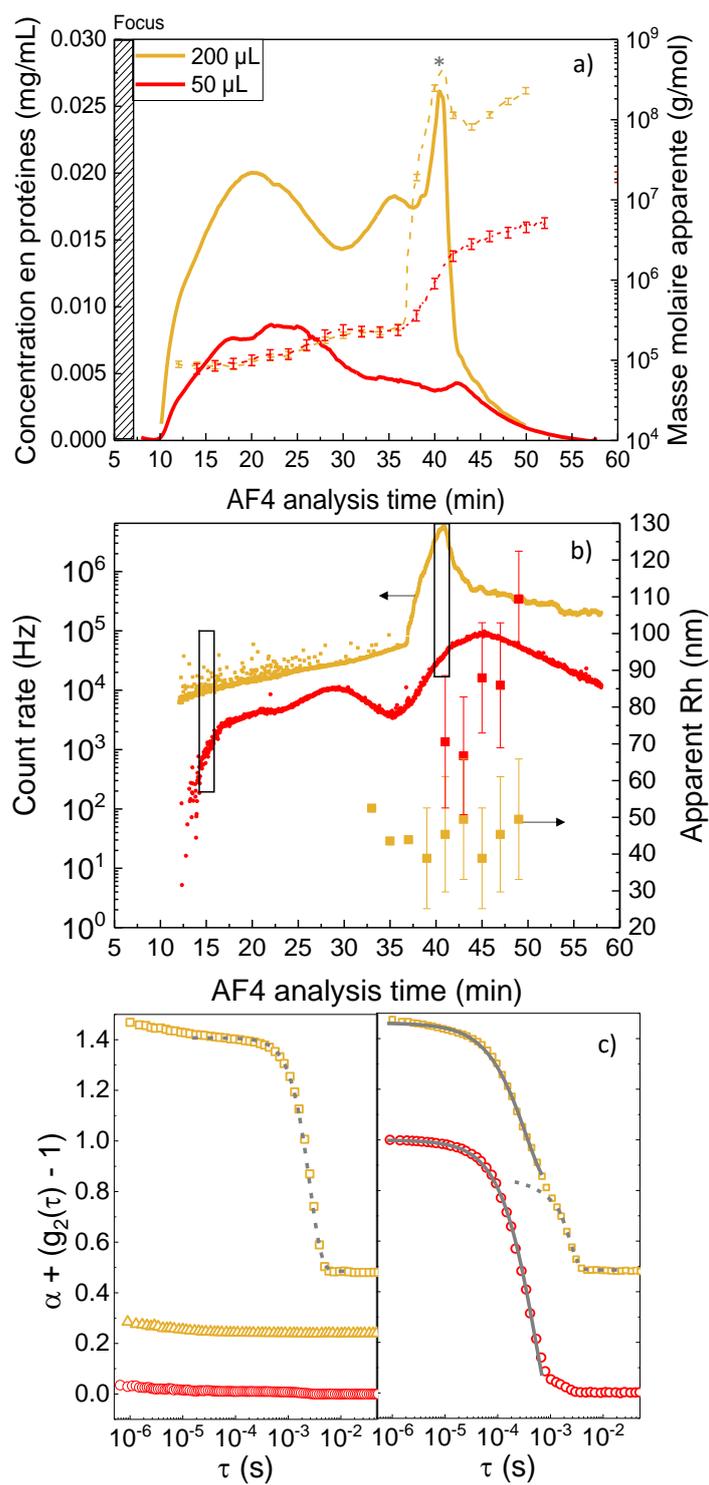

**Figure 5.** AF4 analysis in water/ethanol 50/50 v/v (WE). Two injection volumes are compared: 200 µL (yellow data) and 50 µL (red data) (a) Protein concentration (full line) and apparent molar mass (dashed line) as a function of the AF4 analysis time. The grey star corresponds to the maximum of the MALS signal for the injection volume of 200 µL (b) Dynamic light scattering count rate and hydrodynamic radii as a function of the AF4 analysis time. Error bars on radii represent $1.96\sigma$. The rectangles correspond to the analysis time range over which autocorrelation functions are averaged in (c). (c) Auto-correlation functions measured in-line and averaged over 18-20 min (left) and 42-44 min (right). Autocorrelation functions are shifted vertically with a coefficient ($\alpha = n \times 0.25$, with n an integer) for clarity. Yellow empty triangles in the left plot correspond to the averaged autocorrelation function obtained removing data with a count rate significantly higher than the baseline. Data fitting is performed using cumulant (continuous line) and compressed exponential models (dashed line).

Results are depicted in Figure 5 for two injection volumes: 200 µL and 50 µL. In both cases, the elution step begins at 7 min, and proteins start to be detected at 10 minutes. The 3-minutes difference between the beginning of the protein elution and the end of the focus step suggests that proteins are well focused near the membrane. Indeed, an unproperly relaxed sample would result in an instant sample elution [37] as observed in $P_{0.1}$. Between 10 min and 30 min, a large concentration peak is eluted. Then, proteins continue to be eluted before the apparition of a sharp and intense concentration peak at 40 min, respectively 42 min, for injection volumes of 200 µL, respectively 50 µL. For both injection volumes, between 10 and 20 min, the apparent molar mass measured is constant and equal to $8.5 \times 10^4$ g/mol, it then increases up to $2.0 \times 10^5$ g/mol and plateaus at this value from 30 to 36 minutes. Finally, concomitantly with the final concentration peak, a sharp increase of the apparent molar mass is measured in both cases, but the value reached is very different: up to $4 \times 10^8$ g/mol with 200 µL of injection, and $5 \times 10^6$ g/mol with 50 µL of injection. In-line DLS data are also acquired, but in this solvent, it is difficult to get reliable data at short analysis time due to the very weak intensity of the scattered light (from 10 to 100 times lower in WE than in $P_{0.1}$). Before 32 min, for the injection volume of 50 µL, auto-correlation functions are flat and cannot be exploited (red data in Figure 5c left). With the injection volume of 200 µL, the count rate before 30 min is characterized by the presence of spikes, attributed to strong scattering events due to the sporadic elution of very large objects (aggregates) (Figure 5b). The 2 minutes averaged auto-correlation function measured between 10 and 32 min (blue squares in Figure 5c left) is a compressed exponential. This shows that the dynamics is dominated by the convection of the very large objects (aggregates); (see data at all elution times in Figure SI4 of Supporting Information). The same characteristic time is measured as in conventional AF4 in $P_{0.1}$, which is consistent with the convective motion of large objects since the same flow rate is used (0.6 mL/min). If the spiked data are removed for the calculation of the average auto-correlation function (yellow triangles in Figure 5c left), no exploitable data can be obtained, as for the injection of 50 µL, probably due to the weak contrast of monomeric gluten proteins in WE. After 32 minutes, autocorrelation functions can be fitted with a cumulant model at short time and a compressed exponential model at long time (Figure 5c right). The amplitude of the compressed exponential represents about 30% of the amplitude of the auto-correlation function with 200 µL of injection, while it is negligible with 50 µL. This clearly shows that the co-elution phenomenon found for the injection volume of 200 µL (that is suppressed when decreasing the injection volume down to 50 µL), is mainly due to overloading. This co-elution of large aggregates of unresolved size contributes to the total scattered intensity and invalids the molar mass estimation performed with an injection volume of 200 µL after 35 min of analysis time. For an injection volume of 200 µL, the cumulant fitting of DLS data gives an average size of 45 nm. The polydispersity is low between 32 min and 38 min (polydispersity $\sigma = 2$ nm), and much larger between 40 and 50 min ($\sigma = 15$ nm), suggesting a weaker retention of large objects once the crossflow ramp is stopped. This size is attributed to protein assemblies whose the size is measured a bit larger with an injection volume of 50 µL. With this small injection volume, the hydrodynamic radius (Rh) increases from 70 to 110 nm between 40 and 50 min of analysis time, a range of sizes consistent with the bulk DLS analysis.

In addition, as MALS data are reliable with an injection volume of 50 µL, the protein assemblies molar mass and radius of gyration can be estimated: one measures masses comprised between $1 \times 10^6$ and $5 \times 10^6$ g/mol, and radii of gyration comprised between 50 and 80 nm. For injection volume of 200 µL, eight successive fractions corresponding to time range of 5 minutes of elution were collected, concentrated, and analyzed by SEC. According to the AF4 UV signal, 84% of the injected protein was eluted. As evidenced in the inset of Figure 6a, the mass distribution of the different protein classes eluted coincides well to the injected sample (considering the AF4 recovery of 84±5%), except again the fraction F1 which almost disappears. The recovered protein profiles significantly evolve with AF4 elution. The first collected fraction is mainly composed of α/β, γ-gliadins and the following fractions are progressively enriched in glutenin until 30 min as evidenced by the linear increase of the Glu/Gli ratio from 0.1 to 1.5 (inset of Figure 6b). For analysis times in the range 30-50 min, the protein profiles are very similar whereas the size polydispersity, as evaluated by in line DLS, increases. Hence, the first concentration peak is associated to monomeric proteins that elute progressively according to their increasing molecular size whereas the second concentration peak is associated to monodisperse assemblies; and the last peak would be due to remaining large objects.

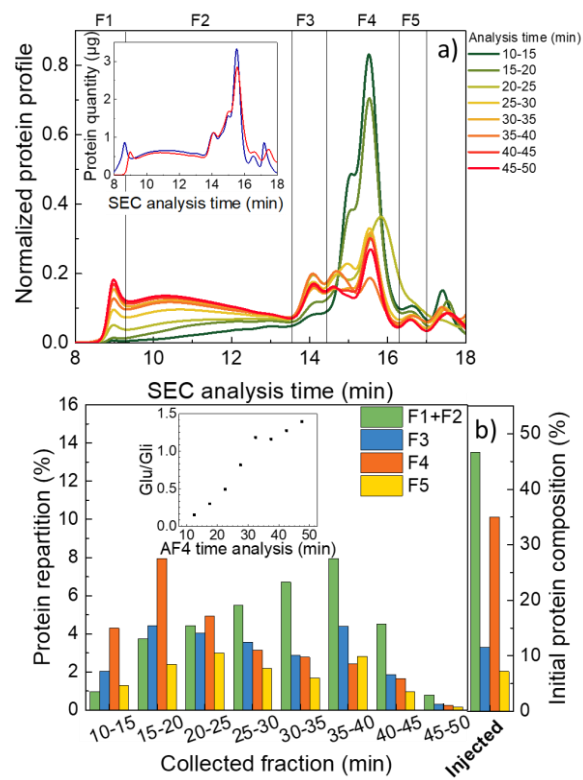

**Figure 6.** SEC analysis of collected samples from AF4 in water/ethanol 50/50 v/v (WE). The injection volume is 200 µL. (a) Normalized protein profiles (the total area is equal to 1) of collected samples. Inset: Comparison of the protein quantities injected multiplied by the AF4 yield (blue) and the quantities collected (red) via the reconstitution of the global SEC profile from the collected profiles. (b) Distribution of the different classes of wheat proteins within collected fractions obtained along elution and comparison with the injected sample composition. Green : glutenin, blue : ω-gliadin, orange : α/β,γ-gliadin and yellow : Albumin/globulin. Inset: evolution of the glutenin/gliadin ratio as a function of the AF4 analysis time.

3.4 Comparison of methods

The AF4 methods, eluents and injection volumes investigated in this study are summarized in Table 2. Generally (excepted for FI-AF4, 50 µL), two concentrations peaks and an intermediate region are evidenced along AF4 analysis time. The average apparent molar mass and the main apparent hydrodynamic radius (excluding co-eluted aggregates) measured for each elution range, are listed. In addition, co-elution of aggregates, probed by the compressed exponential signature of very large scattering objects using in-line DLS and apparent molar masses, is evaluated for each condition. We find that in all cases, when the injection volume is 200 µL, co-elution phenomena occur. This co-elution is suppressed when decreasing the injection volume for FI-AF4 in $P_{0.1}$ and AF4 in WE, while it still occurs using AF4 in $P_{0.1}$. These findings are confirmed by the values of apparent Mw which are consistent with gliadin for the first concentration peak and with glutenin polymers for the second concentration peak, with an injection volume of 50 µL for FI-AF4 in $P_{0.1}$ and AF4 in WE. Interestingly, in these good fractionation conditions, the third concentration peak associated to assemblies is absent in the denaturing solvent $P_{0.1}$ while it is maintained in the low chaotropic solvent WE.

| Fractionation method | | | 1st concentration peak | | | | | Intermediate region | | | | | Final concentration peak | | |
|---|---|---|---|---|---|---|---|---|---|---|---|---|---|---|---|
| Method | Eluent | Injection volume (µL) | Apparent $R_H$ (nm) | Apparent $M_W$ (g/mol) | Glu/Gli | Co-elution of aggregates | | Apparent $R_H$ (nm) | Apparent $M_W$ (g/mol) | Glu/Gli | Co-elution of aggregates | | Apparent $R_H$ (nm) | Apparent $M_W$ (g/mol) | Glu/Gli |
| | | | | | | Intensity | Signature | | | | Intensity | Signature | | | |
| AF4 | $P_{0.1}$ | 200 | 3 | $4.10^5$ | 0.7 | Strong | $M_w$-$R_h$ | 3 | $10^6$ | 0.7 | Strong | $M_w$-$R_h$ | 150 | $3.10^7$ | 1.2 |
| AF4 | $P_{0.1}$ | 50 | 2 | $1.10^5$ | / | Medium | $M_w$-$R_h$ | 3 | $2.10^6$ | / | Medium | $R_h$ | 90 | $6.10^7$ | / |
| FI-AF4 | $P_{0.1}$ | 200 | 3 | $5.10^5$ | 0.4 | Medium | $M_w$-$R_h$ | 3 | $10^6$ | 0.8 | Medium | $M_w$-$R_h$ | 90 | $10^8$ | 1.5 |
| FI-AF4 | $P_{0.1}$ | 50 | 3 | $6.10^4$ | / | Weak | $R_h$ | 3 | $3.10^5$ | / | Weak | $R_h$ | / | / | / |
| AF4 | WE | 200 | / | $8.10^4$ | 0.1 | Medium | $R_h$ | / | $2.10^5$ | 1 | Medium | $R_h$ | 50 | $10^8$ | 1.5 |
| AF4 | WE | 50 | / | $8.10^4$ | / | No | $R_h$ | / | $2.10^5$ | / | No | $R_h$ | 80 | $4.10^6$ | / |

Table 2. Comparison of the different conditions of wheat protein fractionation investigated in the study.

## 3. Discussion

**The delicate analysis of complex samples using Asymmetric Flow Field-Flow Fractionation**

AF4 is a powerful technique to fractionate complex samples showing a wide distribution of sizes, but data analysis can be delicate. In the historical studies, the theory of AF4 was used to deduce the hydrodynamic radius of species from their elution time[13,14,16]. Nevertheless, the theory is only adapted for simple crossflow programs that elute samples free of interaction with the membrane. Hence, a combination of several in-line detectors including UV, refractive index (dRI), MALS, DLS is often used in recent studies in order to better characterize the samples[19,38,39]. However, the analysis of data extracted from the various detectors is not straightforward. The UV signal requires the knowledge of the extinction coefficient of all the eluted species and can be impacted by the light scattering from large objects. In the present study, the absorption associated to the peptide bond, measured at 214 nm, is used to override the variability in the extinction coefficient at 280 nm of the different protein classes (differing in their tyrosine, phenylalanine, and tryptophan contents). In the same vein, both dRI and MALS analysis require the knowledge of the dependence of the index of refraction of the solution with the concentration of species, dn/dC, which also depends on the composition of the eluted species. With the use of surfactants in the elution buffer, protein-surfactant complexes of unknown composition are eluted whereas dn/dC, largely depends on the SDS concentration and can evolve from 0.18 to 0.36 mL/mg[40–42]. In this work we use for dn/dC an average of the experimental measurement of dn/dC in the two solvents ($P_{0.1}$ and WE) for two wheat protein extracts contrasted in terms of composition, one gliadin rich and one glutenin rich. Moreover, as the dRI signal appears very sensitive to variations of pressure associated to the crossflow program we have estimated concentrations from the UV signal. MALS data are exploited with precaution, giving only apparent molecular weight values, as the incertitude on dn/dC values is large, especially in the SDS buffer. In addition, MALS analysis is skewed by co-elution phenomena since conventional theoretical modeling only applies for monodisperse samples. Using in-line dynamic light scattering, hydrodynamic radii can be estimated as long as the diffusive dynamics dominates the signal. Indeed, flow velocity can contribute as well to the intensity fluctuations and auto-correlations functions when Brownian dynamics is slow compared to the translational motion induced by the flow[32]. In these conditions, the flow induced dynamics is ballistic and characterized by compressed exponential auto-correlation functions. The contribution of convection to the autocorrelation function was recently described[43]. Using the experimental parameters used in the present study (i.e solvent viscosity, scattering angle, laser beam size and elution flow rate), the maximum sizes that can be reliably measured ignoring totally the convective contribution are 35, 80 and 50 nm with the conventional AF4 in solvent WE, conventional AF4 in solvent $P_{0.1}$, and SDS frit-inlet methods respectively using equations and criteria defined in [43]. By contrast, the auto-correlation functions are totally dominated by the flow velocity for sizes higher than 465, 1140 and 680 nm in the different methods respectively (See Supporting Information for details of calculations). Hence, a detailed analysis of the shape of the autocorrelation function is crucial to check the role of convection on the measurement and to assess the reliability of the protein sizes extracted. In addition, the analysis of autocorrelations functions can be complexified when co-elution occurs, and a more complex analysis is required.

**Co-elution of wheat protein samples**

Whatever the method used for the fractionation of wheat proteins, when a "large" injection volume is used (200 µL), the co-elution of large objects at short analysis times is evidenced by DLS. In some conditions (in water/ethanol AF4), this contribution dominates the averaged DLS signal whereas it is less important in other conditions (Frit-Inlet). The importance of this contribution depends on several parameters (proportion, size, and contrast of the large objects relatively to the smaller ones) that cannot be disentangled from each other. Nevertheless, the co-elution seems more important in $P_{0.1}$ AF4 as inferred from the MALS signature recorded at the beginning of elution. Indeed, an initial decrease of the apparent molecular weight accompanied by the elution of an important quantity of glutenin polymers, the largest gluten proteins is measured. The co-elution of large objects at the beginning of the fractionation can be characteristic of an insufficient or abnormal focalization before the fractionation. In $P_{0.1}$, the formation of protein/SDS negatively charged complexes increases the electrostatic repulsive interaction. Several studies have demonstrated issues with the use of SDS as AF4 buffer, like the lack of repeatability and lower resolution for the fractionation of wheat protein compared to other surfactants[15]. The authors attribute this to the strong repulsive interaction between the anionic SDS and the regenerated cellulose (RC) membrane which would disrupt the focus step (the RC membrane is indeed negatively charged with a zeta potential of -30 mV at pH=7 [44,45]). Hence, the concentration distribution along the height of the channel would not only result from the size but also from the charge of the objects. Large negatively charged objects could be eluted at shorter time than expected for uncharged objects due to electrostatic repulsion during the focusing/relaxation step. An insufficient focalization can be also due to a too important volume of injection. This is clearly the case for fractionation in WE and FI-AF4 in $P_{0.1}$ for which a smaller volume of injection (50 µL instead of 200 µL) suppresses co-elution phenomena. Nevertheless, in the case of AF4 in $P_{0.1}$, co-elution phenomena are still observed with a small injection volume (50 µL) as evidenced by a decrease of the initial apparent molar mass (See Figure SI2 in Supporting Information). In that case, the co-elution mode could be attributed to the presence of larger aggregates associated to the establishment of a steric mode. The steric elution mode has been widely described in the literature[46–49] especially in the field of nanoparticle fractionation. In this mode, the size of the eluted objects decreases with the elution time, contrary to the normal mode associated to smaller analytes. The presence of large aggregates could result from the concomitant action of the focusing step and the presence of SDS. Indeed, SDS is known to break the non-covalent bonds between protein, but it can also induce protein aggregation in certain conditions, as previously evidenced for lysozyme[50]. Furthermore, the high protein concentration reached close to the membrane during the focalization step can be source of increased interactions, potentially leading to gelation or aggregation. An estimation of protein concentration close to the membrane is given by $C_o = \frac{c_{inj}}{\lambda}$, with $c_{inj}$ the injected protein concentration and $\lambda$ the retention factor[16,51]. The retention factor $\lambda$ is related to the characteristic height of the sample layer in the channel and can be evaluated as $\lambda = \frac{t_o}{6t_r}$, with $t_0$ the void channel time and $t_r$ the retention time of the eluted objects. The void channel time of the conventional AF4 method used here was estimated according to the equation given in ref [44]. Considering a constant crossflow of 1.5 mL/min, we find $t_o$=0.95 min, a value consistent with the experimental evaluation, 1.02 min, obtained for a probe eluted with the AF4 method without focusing step (data not shown). By taking $t_r$=8 min (analysis time of the smallest objects) and $t_r$ = 38 min (analysis time of the largest objects), the retention factor is comprised between 0.004 and 0.019 which corresponds to a maximal protein concentration during the focus step comprised between 100 and 480 g/L. This very high local concentration of proteins would correspond to a semi-dilute regime in which aggregation or gelation of gluten proteins can occur. Indeed, in ethanol/water, the critical gelling concentration of gluten proteins was estimated at 100-200 g/L[53] and an entangled protein network with viscoelastic properties was evidenced above this concentration[22,45]. Hence, the sporadic elution of large scattering objects observed in WE with a volume of injection of 200 µL, could be attributed to a release of particles of gelled proteins detached from the membrane. Concentrated solutions of SDS saturated gluten protein in the buffer $P_{0.1}$ were not previously investigated. We estimate that 1.4 g of SDS is bound per gram of protein[54–56], hence, the SDS concentration during the focus step would be comprised between 140 and 630 g/L. At such high concentration, the protein/SDS complexes could be reorganized to form larger objects.

**Comparison of the different methods for wheat proteins fractionation**

The present study clearly shows that the conventional AF4 method in $P_{0.1}$ fails to fractionate wheat proteins saturated with SDS under the investigated conditions. A strong co-elution is evidenced at short time by dynamic and static light scattering measurements, for the two injection volumes investigated, and the SEC profiles of the collected samples along fractionation show a very weak evolution of the protein composition (Figure 4a). Several studies using AF4 in SDS buffer to fractionate wheat proteins show a similar MALS signature at the beginning of the fractogram, which also suggests the co-elution of large objects[17,18,57,58]. These large objects would result from the concomitant action of SDS and focusing step since they disappear in FI-AF4 with an injection volume of 50 µL.

AF4 fractionation of wheat protein in acid acetic buffer with 0.002% of FL-70 (an alkaline surfactant) using frit-inlet and frit-outlet was found to display a better stability and repeatability than the classic AF4 in the same conditions[59]. Using Frit-AF4 in $P_{0.1}$, the co-elution at the beginning of the fractionation is reduced compared to conventional AF4 for an injection volume of 200 µL and is nearly totally suppressed when the injection volume is 50 µL. The normalized SEC profiles of the collected samples here display a progressive increase of the glutenin/gliadin ratio with elution time, but half of total proteins are eluted before 10 minutes of elution with only a weak enrichment in low molecular weight proteins. In addition to overloading, the limited quality of the fractionation evidenced with 200 µL of injection can be attributed to the fast fractionation of the frit-inlet method that impairs resolution, compared to conventional AF4[58,59]. Furthermore, interestingly, 80 nm sized objects are evidenced with the large injection volume while with 50 µL no assemblies are evidenced, demonstrating the soluble character of these supramolecular assemblies in $P_{0.1}$.

In water/ethanol (WE) the conventional AF4 method shows a good fractionation of the wheat proteins. The ballistic signature measured by DLS with an injection volume of 200 µL suggests co-elution of aggregates (of size of at least hundreds of nanometers) at short elution time but their number would be small as the apparent molar mass of species eluted is consistent with monomeric gliadins that are the main proteins collected at these elution times. In addition, the injection of a smaller sample volume (50 µL), enables the suppression of the ballistic signature associated to aggregates whatever the elution time, allowing a reliable estimation of molar masses. Furthermore, with WE solvent, molar mass analysis is more straightforward, eliminating any uncertainty about the dn/dC value of the SDS-saturated protein. Compared to Morel et al[20] the protocol used here allows a better separation of gliadins due to the constant crossflow at the beginning of the crossflow program. Unfortunately, the Frit-inlet method cannot be tested in WE as the high viscosity of the solvent prevents the use of the high flow rates required for this method and the impact of the focus step cannot be investigated in this solvent.

**Protein assemblies**

In all conditions investigated here, except FI-AF4-$P_{0.1}$-50µL, a concentration peak is measured at long elution time and is associated to protein assemblies with hydrodynamic radii comprised between 50 and 150 nm. The question is whether those assemblies can be compared to the one detected by bulk DLS analysis of the model gluten protein sample. In SEC experiments, owing to their large size, the assemblies are expected to elute at the column void volume (fraction F1). But the SEC fraction F1 accounts for less than 7% of the total proteins, 3 times less than the amount of protein assemblies recovered on AF4 in WE (19%). Furthermore, in all cases the SEC fraction F1 is almost entirely lost during AF4 fractionation. Hence, the protein assemblies eluted on AF4 cannot be assimilated to the SEC fraction F1. We have to admit that the assemblies detected by bulk DLS analysis of the gluten protein samples are not preserved during SEC analysis while they seem to be promoted during AF4, especially in solvent WE and during the sample focusing step. The protein assemblies revealed by AF4 would be dynamic objects obeying a mass action law, vanishing upon an extensive dilution as in FI-AF4 using a small injection volume. They would dissociate when experiencing a shear, as during SEC, while being promoted during sample focusing in AF4. Even more interestingly, the protein composition of these assemblies is identical in the two solvents (WE and $P_{0.1}$) as evidenced by the superposition of the normalized SEC spectra (Figure 7). The assemblies are enriched in ω-gliadins and glutenin polymers, like the glutenin-rich extracts obtained from sequential extraction in dilute HCl[62,63]. The weak association of ω-gliadin with glutenin polymers have been mentioned in several studies[64–66]. Considering the apparent molar mass measured in ethanol/water with an injection volume of 50 μL to avoid any co-elution phenomena, the apparent density of assemblies ($d = \frac{M_w}{4/3 \pi R_H^3 N_a}$) appears comprised between 6 and 7 kg/m$^3$ and the ratio between radius of gyration and their hydrodynamic radius is Rg/Rh=1.12±0.12. This value is typical of branched polymers[67], and is in accordance with our previous studies[20,22]. In addition, their size appears to be dependent on the fractionation conditions in WE, confirming their dynamic character. In the literature, several studies showed the presence of large wheat protein assemblies in several solvents. Protein assemblies with hydrodynamic radius of about 100 nm have been observed in water/ethanol using batch light scattering and AF4[20,22]. Objects with comparable size have been also observed by AF4 in acid acetic buffer[19]. AF4 in SDS buffer $P_{0.1}$ has been extensively applied in view of deciphering the molecular basis of wheat flour breadmaking potential, and the protein fraction exhibiting molecular masses above 2x10$^6$ g/mol was assimilated to "rheologically active polymeric protein"[57]. In this work we show that this AF4 protein fraction cannot be assimilated to exceptionally large glutenin polymers (Mw > 2x10$^6$ g/mol) but involves assemblies of monomeric protein, especially ω-gliadins, with the whole range of glutenin polymers (from $1\times10^5$ to $2\times10^6$ g/mol). Therefore, it cannot be excluded that the breadmaking potential of wheat flour would be rather based on the interaction potential of the glutenin polymers rather than on their intrinsic molecular size, as claimed since a long time[68].

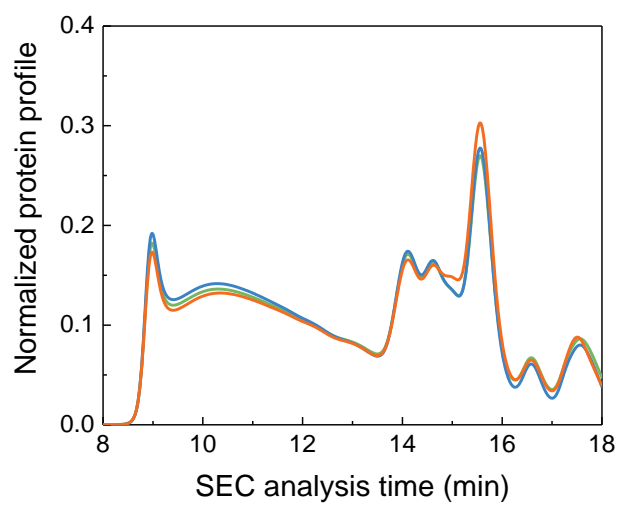

**Figure 7.** Comparison of normalized SEC protein profiles obtained from fractions collected at the maximum MALS signals (grey star in Figures 3a and 5a): (green) 45-50 min fraction from AF4 in water/ethanol 50/50 v/v (WE), (blue) 30-34 min fraction from Frit-AF4 in 0.1M phosphate + 0.1% SDS ($P_{0.1}$) and (orange) 45-50 min fraction from AF4 in $P_{0.1}$.

## 4. Conclusion

We have shown that the fractionation of wheat proteins is delicate due to their tendency to self-assemble. SDS enables the total solubilization of the model gluten protein extract but the fractionation by Flow Field technique in this solvent is often characterized by co-elution phenomena, induced by the AF4 focusing step and overloading. We have evidenced one condition (FI-AF4-50μL), to fractionate wheat proteins as individual proteins, without co-elution of large objects, using the denaturing solvent including SDS. In this condition, the supramolecular assemblies evidenced in bulk DLS are totally suppressed, showing their soluble character. In water/ethanol, AF4 displays a great capacity to fractionate wheat proteins. In this condition supramolecular assemblies enriched in glutenin and ω-gliadin are detected. Interestingly, protein assemblies of the same composition were also measured at long elution time in the SDS buffer in conditions inducing co-elution, showing that the highest molar mass objects probed in this solvent are also supramolecular and do not correspond to isolated glutenin polymers as previously postulated in the literature[16,19]. In addition, their size depends on analysis conditions, especially because of their dynamic character. The dynamic nature of wheat protein supramolecular assemblies would certainly require further investigations.

**Supporting Information:**

Auto-correlation functions measured in bulk for the model gluten extract dispersed in both solvents. Protein concentration and molar mass measured with AF4 methods in $P_{0.1}$ using injection volumes of 50 and 200 μL. Experimental in-line DLS autocorrelation functions together with the fit of data for all data not shown in the main manuscript. Recovery of all AF4 injections.

**Authorship contribution statement:**

A. Urbes: Investigation, Visualization, Writing original draft. M.-H. Morel: Conceptualization, Resources, Investigation, Visualization, Writing, Supervision. F. Violleau: Project administration, Resources, Funding acquisition. L. Ramos: Conceptualization, Review. A. Banc: Conceptualization, Investigation, Visualization, Writing - review & editing, Funding acquisition, Project administration, Supervision.

**Acknowledgments:**

Région Occitanie and Ecole d'Ingénieur Purpan are acknowledged for the funding of the phD grant of A. Urbes. The work was also financially supported by the French National Agency through the young researcher project entitled Elastobio, grant: ANR-18-CE06-0012-01.

**Abbreviations:** AF4, Asymmetric Flow Field Flow Fractionation; SDS, Sodium Dodecyl Sulfate; DLS, Dynamic Light Scattering; MALS, Multi Angle Light Scattering; SEC, Size Exclusion Chromatography; UV, Ultraviolet; dRI, differential Refractive Index; Mw, Molecular Weight; SDS-PAGE Sodium Dodecyl Sulfate – PolyAcrylamide Gel Electrophoresis

# Supporting information

1. **Characterization of the model protein extract – Batch DLS**

The auto-correlation functions of the model protein isolate dispersed in solvent $P_1$ and WE are measured at different scattering angles as described in the main manuscript. The autocorrelation functions are fitted with a double exponential model assuming a bidisperse suspension (Figure SI1 a and b). The slow and fast decay rates obtained by the fit are plotted as a function of the square of the scattering vector q (Figure SI1 c and d). The fast decays increase linearly as a function of $q^2$ in the whole range of q and their slope allows an estimation of the small objects present in the solution: 20 nm and 22 nm, in $P_1$ and WE respectively. By contrast, the slow decay rate no longer evolves linearly as a function of $q^2$ above $q^2 = 4.10^{14}$ $m^{-2}$, which may be characteristic of the object's internal motion. The slope of the linear regime between 0 and $4.10^{14}$ $m^{-2}$ allows one to estimate a size of 245 nm for the model protein extract in $P_1$ and 129 nm in WE.

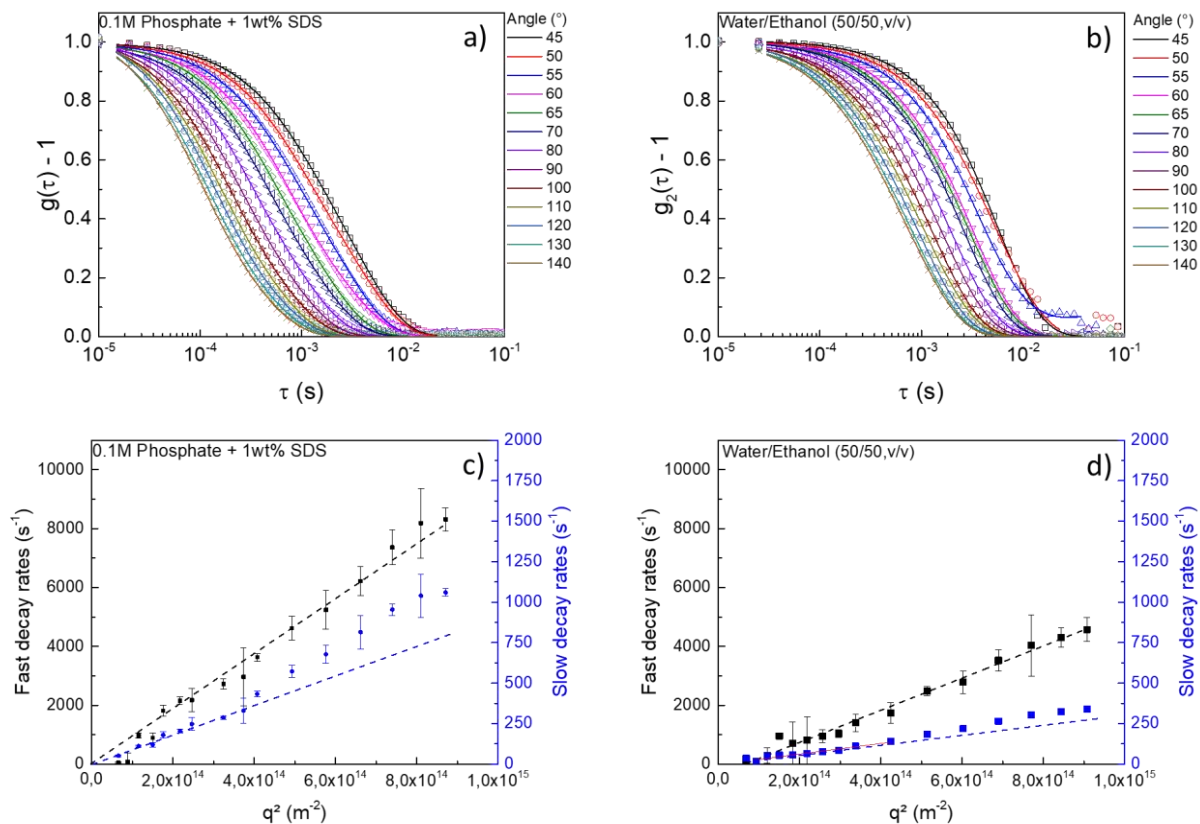

**Figure SI 1.** Auto-correlation functions measured at different scattering angles as indicated in the legend, using batch DLS for the model protein extract dispersed in 0.1 M phosphate buffer, pH 6.8, 1% SDS ($P_1$) (a) and in water/ethanol 50/50 v/v (WE) (b). Evolution of the fast and slow decay rates as function of q² for the sample dispersed in $P_1$ (c) and in WE (d).

2. **Fractogram of the model protein extract at different injection volumes in $P_{0.1}$**

Figure SI 2 displays the protein concentration and the apparent molar mass as a function of the AF4 analysis time in $P_{0.1}$ for injection volumes of 50 μL and 200 μL (Figure SI 2, black and red lines, respectively). For both injection volumes, AF4 in $P_{0.1}$ is characterized by an initial decrease of apparent molar masses which is the signature of the steric mode. At elution times corresponding to the first concentration peak, for both methods, the apparent molar mass is higher with the large volume of sample injected (Figure SI2 a), that evidence overloading when the large injection volume is used. In addition, the apparent molar masses evolutions are different with the two injection volumes. Besides, a clear different protein concentration profile is measured with the two injection volumes. The positions of concentration peaks are roughly identical, but their relative amplitudes are significantly different. This could be due to a concentration dependence of the assemblies, which will be the object of future investigations.

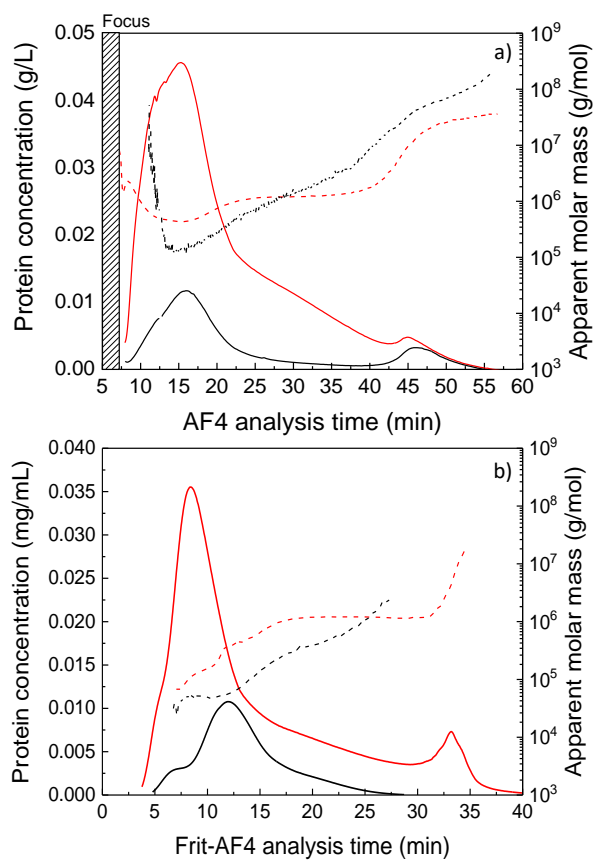

**Figure SI 2.** Protein concentration (full lines) and apparent molar mass (dashed lines) as a function of the analysis time for the model protein extract injected at 2 mg/mL with injection volumes of 50 µL (black) and 200 µL (red). (a) AF4 in $P_{0.1}$ and (b) Frit-AF4 in $P_{0.1}$.

3. **Fits of in-line DLS autocorrelation functions measured in $A_{0.1}$ using an injection volume of 50μL.**

Figure SI3 displays the autocorrelation functions averaged each two minutes measured by the in-line DLS detector along AF4 (a) and Frit-AF4 (b) fractionation in solvent $P_{0.1}$ for an injection volume of 50 μL (a). The autocorrelation functions are fitted with a double exponential (equation 1 of the main manuscript) for all measurements, and results of fitting are given in Table SI 1. The long decay time, around $10^{-3}$ s (associated to apparent $R_H$ comprised between 200 and 350 nm in table SI 1), is associated to the convective motion of large scattering objects of unresolved size. Its relative contribution is lowered compared to the equivalent AF4 experiment performed with an injection volume of 200 μl (see figure 3 of the manuscript). In Frit-AF4, using an injection volume of 50 μL the long decay time is even negligible and indicates that co-elution is nearly suppressed. In A4F-50μL, until 30 min of analysis time, the short decay time is attributed to a hydrodynamic size of about 2 nm. Between 30 and 40 minutes, the short decay time gives a size of 1-2 nm, which is very small for proteins. Nevertheless, this surprising result coincides with the noisy appearance of auto-correlations functions and the negligible protein concentration. Beyond, 3 nm sized objects are found in equilibrium with 80-150 nm assemblies. By contrast, in Fl-AF4, the same size of about 2 nm is obtained whatever the analysis time and no assemblies are evidenced.

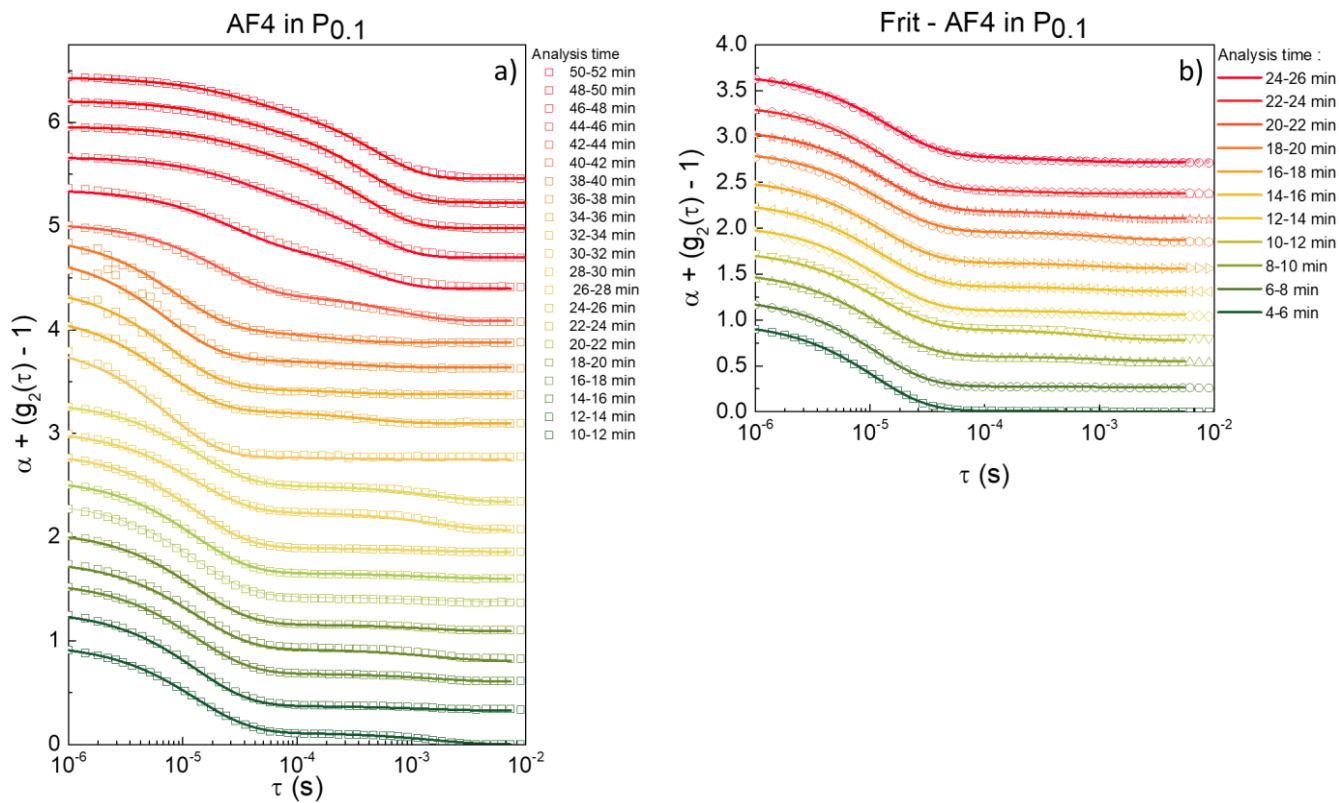

**Figure SI 3.** Auto-correlation functions measured in-line for AF4 (a) and Frit-AF4 (b) in $P_{0.1}$ using an injection volume of 50 µL. Data fitting is performed using a double exponential model. Data are shifted vertically for clarity.

AF4 - $P_{0.1}$ - 50 µL

| Analysis time | 10-12 min | 12-14 min | 14-16 min | 16-18 min | 18-20 min | 20-22 min | 22-24 min | 24-26 min | 26-28 min | 28-30 min | 30-32 min | 32-34 min | 34-36 min | 36-38 min | 38-40 min | 40-42 min | 42-44 min | 44-46 min | 46-48 min | 48-50 min | 50-52 min |
|---|---|---|---|---|---|---|---|---|---|---|---|---|---|---|---|---|---|---|---|---|---|
| $A_f$ | 0.65 | 0.77 | 0.72 | 0.65 | 0.73 | 0.77 | 0.75 | 0.79 | 0.56 | 0.58 | 0.88 | 0.68 | 0.76 | 0.72 | 0.66 | 0.54 | 0.67 | 0.70 | 0.65 | 0.73 | 0.77 |
| $A_s$ | 0.35 | 0.23 | 0.28 | 0.34 | 0.26 | 0.21 | 0.25 | 0.20 | 0.44 | 0.42 | 0.12 | 0.32 | 0.24 | 0.28 | 0.34 | 0.46 | 0.33 | 0.28 | 0.34 | 0.26 | 0.23 |
| $R_{hf}$ (nm) | 2 | 3 | 2 | 2 | 2 | 2 | 2 | 2 | 2 | 2 | 1 | 1 | 1 | 1 | 1 | 3 | 3 | 3 | 3 | 4 | 3 |
| $R_{hs}$ (nm) | 326 | 269 | 323 | 326 | 290 | 269 | 274 | 251 | 341 | 345 | 351 | 279 | 351 | 262 | 236 | 155 | 84 | 83 | 150 | 178 | 156 |

Frit-AF4 - $P_{0.1}$ - 50 µL

| Analysis time | 4-6 min | 6-8 min | 8-10 min | 10-12 min | 12-14 min | 14-16 min | 16-18 min | 18-20 min | 20-22 min | 22-24 min | 24-26 min |
|---|---|---|---|---|---|---|---|---|---|---|---|
| $A_f$ | 0.70 | 0.88 | 0.89 | 0.91 | 0.78 | 0.75 | 0.73 | 0.84 | 0.88 | 0.93 | 0.78 |
| $A_s$ | 0.30 | 0.12 | 0.11 | 0.07 | 0.22 | 0.24 | 0.26 | 0.16 | 0.12 | 0.17 | 0.26 |
| $R_{hf}$ (nm) | 2 | 2 | 2 | 2 | 2 | 2 | 2 | 2 | 2 | 2 | 2 |
| $R_{hs}$ (nm) | 7 | 8 | 211 | 202 | 13 | 11 | 9 | 248 | 192 | 54 | 84 |

**Table SI 1:** Fitting parameters of the auto-correlation functions measured along elution of AF4 (a) and Frit-AF4 (b) in $P_{0.1}$ using an injection volume of 50 µL.

The apparent sizes of assemblies eluted during the final concentration peak, for different $P_{0.1}$ conditions, are summarized in Table SI 2.

| | AF4 in $P_{0,1}$ | | | |
|---|---|---|---|---|
| **Analysis time (min)** | **44-46** | **46-48** | **48-50** | **50-52** |
| Average radius for $V_{inj}$ = 50 µL (nm) | 85 ± 9 | 82 ± 10 | 85 ± 6 | 99 ± 5 |
| Average radius for $V_{inj}$ = 200 µL (nm) | 83 ± 2 | 150 ± 13 | 178 ± 15 | 156 ± 16 |
| | Frit-AF4 in $P_{0,1}$ | | | |
| **Analysis time (min)** | **28-30** | **30-32** | **32-34** | **34-36** |
| Average radius for $V_{inj}$ = 200 µL (nm) | 115 ± 2 | 73 ± 7 | 102 ± 4 | 180 ± 15 |

**Table SI 2** Apparent hydrodynamic radii of assemblies obtained using the short time of the double exponential model as function of the AF4 and Frit-AF4 analysis time in $P_{0.1}$.

### 4. Fits of in-line DLS autocorrelation functions measured by AF4 in WE.

Figure SI4 displays the autocorrelation functions (in-line DLS detector) averaged on two minutes intervals along AF4 fractionation in solvent WE for the two injection volumes: 50 µL (a) and 200 µL (b). The autocorrelation functions measured with an injection volume of 200 µL are fitted with a compressed exponential for analysis time comprised between 10 and 30 min. The autocorrelation functions measured between 30 and 50 min are fitted with a cumulant function at short decay times and with a compressed exponential function at long decay times. The fitted compression factor (β) is comprised between 1.98 and 2.30 for all autocorrelation functions. With an injection volume of 50 µl, the autocorrelation functions cannot be exploited before 36 min of analysis time and beyond are correctly fitted with a cumulant model. The apparent sizes estimated from the cumulant model are summarized in table SI 3.

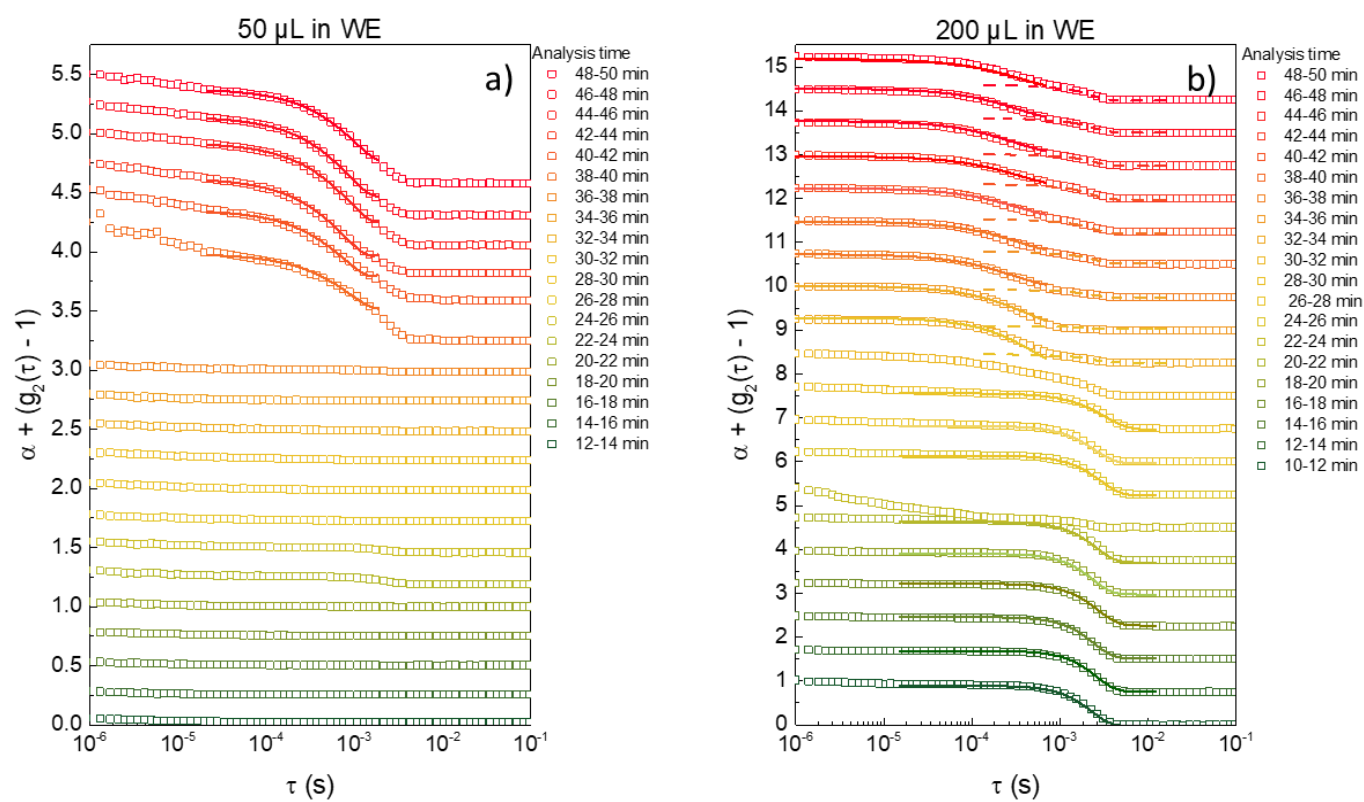

**Figure SI 4.** Auto-correlation functions measured in-line for AF4 in WE performed with injection volumes of 50μL (a) and 200μL (b). Data fitting is performed using a compressed exponential in (a), a cumulant model at short decay time and a compressed exponential at long decay time in (b). Data are shifted vertically for clarity.

| Analysis time (min) | 38-40 | 40-42 | 42-44 | 44-46 | 46-48 | 48-50 |
|---|---|---|---|---|---|---|
| Average radius for $V_{inj}$ = 50 μL (nm) | 91 ± 19 | 70 ± 18 | 67 ± 16 | 88 ± 15 | 86 ± 17 | 109 ± 13 |
| Average radius for $V_{inj}$ = 200 μL (nm) | 44 ± 2 | 39 ± 13 | 45 ± 15 | 50 ± 16 | 39 ± 13 | 45 ±1 5 |

**Table SI 3.** Apparent hydrodynamic radii of the assemblies obtained using the cumulant model as function of the AF4 analysis time for the two injection volumes in WE.

## 5. A4F recovery of the different injections

| AF4 in $P_{0.1}$ | | Frit-AF4 in $P_{0.1}$ | | AF4 in WE | |
|---|---|---|---|---|---|
| For n analysis (n=6) | | | | | |
| 50 µL | 200 µL | 50 µL | 200 µL | 50 µL | 200 µL |
| 83 ± 3% | 87 ± 5% | 8 ± 6 % | 81 ± 6 % | 91 ± 4 % | 84 ± 5% |

Table SI summarizes the AF4 recovery of the different injections. The AF4 recovery (in %) is obtained by dividing the amount of protein eluted (calculated from the AF4 UV signal) to the total amount of protein injected.

| AF4 in $P_{0.1}$ | | Frit-AF4 in $P_{0.1}$ | | AF4 in WE | |
|---|---|---|---|---|---|
| For n analysis (n=6) | | | | | |
| 50 µL | 200 µL | 50 µL | 200 µL | 50 µL | 200 µL |
| 83 ± 3% | 87 ± 5% | 8 ± 6 % | 81 ± 6 % | 91 ± 4 % | 84 ± 5% |

**Table SI 4.** AF4 recoveries

## 6. Regimes of DLS analysis under flow

According to Chowdhury et al.[1], the total displacement of monodisperse Brownian particles under flow can be approximated by a sum of Brownian motion and linear flow, leading to an auto-correlation function in the form:

$$g_2(\tau) = B(1 + \beta \exp(-2\Gamma\tau) \exp\left(-\frac{v^2\tau^2}{\omega^2}\right)) \text{ equation 1}$$

with $\Gamma$ the Brownian decay constant, $v$ the flow velocity, $\omega$ the beam radius, $B$ the baseline ($B \approx 1$) and $\beta$ the spatial coherence factor. The first exponential term is associated to the stochastic Brownian motion whereas the second exponential term, that is a compressed exponential, is associated to the linear flow, or transit, of these particles in the scattering volume. Experimental auto-correlation functions measured under flow can be modelled by this mixed model considering the two contributions (Equation 1). However, if the characteristic decay times associated to the Brownian motion and the linear flow are significantly different, the decay of the auto-correlation function is dominated by the fastest process. Torquato et al.[2] defined 4 regimes of fitting strategies required to analyze DLS data under flow. For a given linear flow velocity, ν, the maximum size hydrodynamic radius, $R_{H\ Max}$, that can be defined using a given strategy is:

$$R_{H\ Max} \sim \alpha \frac{q^2 \omega kT}{3\pi \upsilon \eta}$$

With $\alpha$ =0.142 for the "quiescent regime" in which the transit contribution can be ignored.

With $\alpha$=0.536 for the "transit approximation regime" in which the mixed model (equation 1), can be correctly applied.

With $\alpha$=1.98 for the "model deviation regime" for which the mixed model no longer correctly describe data.

For sizes above, we enter in the "breakdown regime" in which autocorrelation functions are totally dominated by the transit contribution and no size can be determined.

The limits of the different regimes are calculated in our experimental AF4 conditions using the beam radius ω=2.10$^{-5}$ m, the scattering vector q, the eluent viscosity $\eta$, the flow velocity $\upsilon=\frac{Q}{\pi r^2}$ with Q the outlet flow (in m$^3$/s) and r the radius of the light scattering measuring cell (r=1.2mm) and summarized in Table SI 5.

|               | Quiescent regime | Transit regime | Deviation regime |
|---------------|------------------|----------------|------------------|
| WE            | 33               | 126            | 465              |
| AF4 in $P_{0.1}$    | 82               | 307            | 1136             |
| Frit-AF4 in $P_{0.1}$ | 49             | 185            | 681              |

**Table SI 5.** Limits of the different fitting strategies regimes for the fractionation methods used in the study. $R_{H\,Max}$ values (given in nm) correspond to maximum hydrodynamic radii associated to the different regimes.